\documentstyle[12pt]{article}
\input epsf

\newlength{\extraspace}
\newlength{\extraspaces}
\newcommand{\be}{\begin{equation}
\addtolength{\abovedisplayskip}{\extraspaces}
\addtolength{\belowdisplayskip}{\extraspaces}
\addtolength{\abovedisplayshortskip}{\extraspace}
\addtolength{\belowdisplayshortskip}{\extraspace}}
\newcommand{\ee}{\end{equation}}

\newcommand{\bq}{\begin{eqnarray}
\addtolength{\abovedisplayskip}{\extraspaces}
\addtolength{\belowdisplayskip}{\extraspaces}
\addtolength{\abovedisplayshortskip}{\extraspace}
\addtolength{\belowdisplayshortskip}{\extraspace}}
\newcommand{\eq}{\end{eqnarray}}

\newcommand{\newsection}[1]{
\vspace{15mm}
\pagebreak[3]
\addtocounter{section}{1}
\setcounter{equation}{0}
\setcounter{subsection}{0}
\setcounter{footnote}{0}
\begin{flushleft}
{\large\bf \thesection. #1}
\end{flushleft}
\nopagebreak
\medskip
\nopagebreak}

\begin{document}
\hbox{}
\nopagebreak
\vspace{-3cm}
\addtolength{\baselineskip}{.8mm}
\baselineskip=24pt
\begin{flushright}
{\sc OUTP}- 94-39  P\\
hep-th@xxx/9412232 \\
 December  1994
\end{flushright}
\vspace{0.4in}
\begin{center}
{\Large  Black Hole  Spectrum, Horison Quantization and
All That:
 $(2+1)$-Dimensional Example .}\\
\vspace{0.2in}

{\large Ian I. Kogan}
\footnote{email: kogan@thphys.ox.ac.uk}
\footnote{ On  leave of absence
from ITEP,
 B.Cheremyshkinskaya 25,  Moscow, 117259, Russia}\\
{\it  Theoretical
 Physics \\
1 Keble Road, Oxford, OX1 3NP, UK}
\\
\vspace{0.3in}

{\sc \Large  Abstract}
\end{center}

\noindent

We discuss the possibility  that quantum black holes
 have discrete mass spectrum. Different arguments
 leading to this conclusion are  considered, particularly
 the  decoupling between left and right sectors in string theory -
 the so-called heterotic principle.
 The case  of a $2+1$ dimesnional black holes
 is considered  as an argument in favour of this argument. The
  possible connection between membrane model of the black hole horison
 and topological membrane is briefly discussed.

\vfill

\newpage
\newsection{Introduction.}

 Quantum black holes   play the same role now as
 black body radiation did a century ago, creating  problems and
 paradoxes which the future theory (presumably of the twenty-first century)
 will resolve.  The fact
 that the most intriguing  process occuring with quantum black holes,
  Hawking quantum evaporation \cite{hawking},
  describes the outcoming radiation as having
   (almost) the  black body spectrum,  supports in an amusing way  a
  continuity of tradition in theoretical physics. Using this as a
 guideline one   can ask the question
 whether the   quantum black hole is an analog of another cornerstone of
 twentieth-century quantum physics - the hydrogen atom. Do
 quantum black holes have a discrete  spectrum and if yes -
 may it be  that the Hawking evaporation is the transition
  process  between the discrete states ?

The idea that black holes may have a discrete spectrum was first
 proposed by Bekenstein in 1974 \cite{bekenstein} who used an analogy
 between a horison area $A$ for Kerr black hole,
  proportional to the squared irreducible mass $M^{2}_{ir}
 = A/16\pi$,  and an action integral
 $\oint pdq$ of a periodic mechanical system. This analogy
 was based on a fact that the  irreducible mass behaves as an
 adiabatic invariant, i.e. remains unchanged in reversible processes
  (existing only for Kerr black hole with non-zero angular momentum,
 see  \cite{bekenstein} for details). Using this analogy and Bohr-Sommerfeld
 quantization condition Bekenstein  obtained the discrete  spectrum
\bq
M^{2}_{ir}
  \sim  M_{p}^{2} n
\eq
 where $M_{p}$ is the Planck mass and  $n$ is an
integer.
 Later,  in 1986  V.Mukhanov \cite{mukhanov} and
author\cite{kogan1}
\footnote{Unfortunately at that time I was unaware
 about Bekenstein paper \cite{bekenstein} as well as now
 have no information   about any paper
  discussing  this subject between 1974 and 1986}
 independenly revived this idea   using completely different
arguments, which will be briefly discussed later.
   The last year  this problem attracted  more attention and
  has been  discussed  in  several
 interesting   papers \cite{gb} - \cite{lomag},   where the discrete
 spectrum was   derived  using new  ideas.  Let us
 also note that quantization of the area operator in quantum gravity
 has been obtained in \cite{ars}, \cite{smolin} using the loop
 representation.

 Recently the new class of black holes in three-dimensional spacetime
with a negative cosmological constant was considered  by Banados,
 Teitelboim and Zanelli \cite{btz} (see also detailed paper
 \cite{bhtz}). The particular case of  charged black holes  in the
 limit of zero cosmological constant as well as black hole solutions
in  $U(1)$ topologically massive gauge theory (i.e.  $2+1$ gauge
 theory with a Chern-Simons term)
   has been considered at the  same
 time in \cite{kogan2}.

  In two recent papers Maggiore \cite{maggiore2} and Carlip
\cite{carlip} studied the entropy of  neutral $2+1$ black holes
 \cite{btz}, \cite{bhtz}, \cite{ct}. In \cite{maggiore2}
 the membrane approach, which has been previously applied to $3+1$
 black holes \cite{maggiore1}, \cite{lomag}, reproduced the correct
value of
entropy. The Schroedinger equation for the horizon wave function
  has been written,  which leads to the discrete spectrum (however
 the explicit form of the spectrum has not been discussed there).
  In \cite{carlip} the similar picture was obtained, where the
 ``membrane'' degrees of freedom  at the horison were nothing but the
 gauge degrees of freedom arising due to the breaking of gauge
invariance of $2+1$-dimensional gravity due to the presence of horizon.
 The entropy was obtained as the logarithm of the number of these
 states.   But in obtaining this the relation between the horizon
 radius $r_{+}$ and  a number operator $N$ with integer spectrum
 has been obtained
\bq
 N = \left(\frac{r_+}{4G}\right)^{2}
\label{spectrum2+1}
\eq
which means the quantum spectrum for mass $M = r_{+}^{2}/l^{2}$, where
 $l$ is an  inverse cosmological constant, takes the following form
\bq
M = \left(\frac{4G}{l}\right)^{2} N
\eq
Let us note that it  is proportional to $N$ and
 not to  $\sqrt{N}$ as in $3+1$ case.

The aim of this paper is to discuss this spectrum  as well as spectrum
arising  for charged black holes and  find what approach used
 previously  lead to this spectrum. In the next section we shall
 discuss arguments used in \cite{mukhanov} and \cite{kogan1}. Than we
 shall remind some basic facts about $2+1$ black holes and briefly
 discuss the Carlip approach \cite{carlip} leading, as we shall
demonstrate, to the spectrum (\ref{spectrum2+1}). Generalization
 of this   quantization condition in case of  black holes discussed
 in  \cite{kogan2} will be given.  Because entropy and action both
 proportional to $r_{+} \sim \sqrt{N}$ \cite{btz} it is not so
 easy to get this spectrum from  Bohr-Sommerfeld
 quantization. However this  spectrum is in agreement with
string-inspired arguments presented in \cite{kogan1}. In conclusion
 we shall discuss possible connection between these arguments and
 Maggiore membrane approach \cite{maggiore1},\cite{lomag},
 \cite{maggiore2}.

\newsection{Heuristic arguments for black hole quantization.}
 Now we shall consider arguments presented in \cite{mukhanov} and
 \cite{kogan1} originally in the case of $3+1$-dimensional
 black holes and discuss how they can be, in principle, generalized.
 Let us note that both these arguments, contrary to the original
 Bekenstein idea,  relied upon the existence of black hole temperature
  and Hawking evaporation,
 which  became known only  after Bekenstein proposal \cite{bekenstein}
\subsection{Thermodynamics and Hawking evaporation.}
The first argument \cite{mukhanov} (see also  \cite{gb}) used the
 first law of thermodynamics for black holes \cite{bekenstein2}
 \bq
 \delta M = \frac{1}{4} T \delta A + \Omega \delta J +
\Phi \delta Q
\eq
where $T, \Omega$ and $\Phi$ are the temperature, rotational frequency
 and electric potential of the black hole with mass $M$, surface area
$A$, angular momentum $J$ and charge $Q$,
  and an assumption
  that   Hawking   evaporation can be described as the  result  of
 a spontaneous transition between discrete levels of black hole and
 it was assumed that  transitions occur between   nearest level, let
 say $n$ and $n-1$.
 Then  for radiation mode characterized by a frequency  $\omega$,
a charge $e$ and azimutal quantum number $m$ one has
\bq
\omega - e\Phi - m\Omega = \alpha T
\eq
 where $\alpha =ln
2$ if one assumes \cite{mukhanov}, \cite{gb} that during each
transition the emitted quantum carries a minimum quantum of
information - one bit.
   Because $\omega = M_{n}-M_{n-1} = \delta M$
 one can see immdeiately that $\delta A = A_{n} - A_{n-1} = 4\alpha$
 and  $A_{n} = 4\alpha n$  which predicts the mass spectrum of
 the Schwarzchild black hole ($\Omega = \Phi = 0$)
\bq
M_{n} \sim M_{p} \sqrt{n}
\eq

Let us note that we got this spectrum making the assumption that
 $\omega = M_{n}-M_{n-1} \sim T$. However one can imagine the
situation  when transitions between all levels are of the same order
 of magnitude and in this case the radiation temperature, i.e. the
 characterisitc energy  emitted, may be much larger than the distance
 between initial energy level $M_{n}$ and the closest one $M_{n-1}$.
 For example of somebody gets the mass spectrum
  $M_{n} \sim M_{p} n$ and $T_{n} \sim M_{p}\sqrt{n}$
 (as we shall see later  this is true for $2+1$ dimensional black
holes),
 i.e.  at large $n$ one has $T >> \delta M =   M_{n}-M_{n-1}$  and  to
 avoid the paradox one must assume that there  are
  unsuppressed transition matrix elements $<n|m>$ for all
 $|n-m| \leq \sqrt{n}$. Then the average  energy which may be
 emitted  will be  by order of the maximal possible level splitting,
 i.e.  $M_{n+\sqrt{n}} - M_{n} \sim M_{p}\sqrt{n}
 \sim T$ and one can try  to reproduce the Hawking spectrum simply
 changing the pattern of the transition matrix elements.

\subsection{String winding modes and left-right decoupling.}

In \cite{kogan1} another approach to the quantization was suggested
 based on a picture of a test string in a black hole gravitational
 background
\bq
S = \frac{1}{2\pi} \int d^2\xi G_{\mu\nu}(x) \partial_{a}x^{\mu}
 \partial_{a}x^{\nu}
\label{targetbh}
\eq
where the  Schwarrzschild metric (after analytical continuation into
 the Euclidean region) takes the form
\bq
ds^{2} = G_{\mu\nu}(x) dx^{\mu}dx^{\nu} =
 \left(1- \frac{r_{+}}{r}\right) dt^2 +
 \left(1- \frac{r_{+}}{r}\right)^{-1} dr^{2} + r^{2}\left(d\theta^{2}
+ \sin^{2}\theta d\phi^{2}\right)
\eq
where $r_{+}$ is a horison radius  proportional to the black hole mass
$M = M_{p}^{2}r_{+}$.
This metric describes a  regular manifold with the
 topology $R^{2}\times S^{2}$ provided $r \geq r_{+}$ and imaginary
 time $t$ is an angular variable with periodicity $4\pi r_{+}$, which
 means that we have the temperature $T = 1/4\pi r_{+} = M_{p}^2/4\pi M$.
 This periodicity can be seen by making the substitution $r = r_{+}
 + y^2/4r_{+}$ in the limit $r \rightarrow r_{+}$, then metric becomes
 $ds^2 = dy^2 + y^2 d(t/2r_{+})^{2} + r_{+}^{2}d\Omega^2$ and to
 avoid the conical singularity at $y = 0$ one has to make $ t/2r_{+}$
 angular variable with  a period $2\pi$.

  Let us examine the equations of motion which follows from
 (\ref{targetbh})
\bq
\partial_{z}\partial_{\bar{z}} x^{\mu} + \Gamma^{\mu}_{\nu\rho}(x)
\partial_{z}x^{\nu}\partial_{\bar{z}} x^{\rho} = 0
\eq
where $z = \sigma + \tau$ and  $\bar{z} = \sigma - \tau$.
 It is easy to see that there are exact  left- or right-moving solutions
of the form
\bq
x_{L} = x(z) = x(\tau+\sigma), ~~~x_{R} = x(\bar{z}) = x(\tau -\sigma)
\eq
 Now let us  ask the question if these chiral solutions exist for
 general background, i.e.  for general horison radius $r_{+}$ which,
 in our case, determines the metric. Expanding chiral (let say left)
 solution  in oscillators
\bq
x_{L}(\tau+\sigma) = x_{L} + p_{L}(\tau+\sigma) + \frac{i}{2}
\sum_{n\neq 0}\frac{1}{n} a_{n} \exp\left[2in(\tau+\sigma)\right]
\eq
If one demands the absence of the right mode we are dealing with the
 winding mode where  $p_{L}$ is the winding number.  One can ask the
 question how it is possible to consider only left modes and have no
 right mode at all.   Generally speaking   both sector - left and right
 - exist in a given background. However in case of Eucledian metric
 of black hole one  must remember that it is only the part of the space-time
  with $r> r_{+}$  which  has been  analytically continued. One can
 imagine  the situation (not in string theory, but in topological
 membrane where left- and right  world sheets are independent) when left
 sector is above the horison and right one is beyond. In this case
  after analytical continuation we indeed have  only left movers.

   It is easy to see that it  is impossible to have nonzero
 $p_{L}$ in general case -
 because we are dealing with closed strings $x^{\mu}_{L}(\tau) =
 x^{\mu}_{L}(\tau + 2\pi)$ which excludes any nonzero $p^{\mu}_{L}$.
 However this is not  correct for  $x^{0} = t$ mode. Because
 of periodicity it is possible to have  $x^{0}_{L}(\tau) =
 x^{0}_{L}(\tau + 2\pi) + 4\pi m r_{+}$ where $m$ is some integer -
 winding number. This gives us
 $p^{0}_{L} \sim m r_{+}$. However because $p^{0}_{L}$ is the momentum
conjugate
 to the periodic variable it spectrum must be $p^{0}_{L} \sim n / r_{+}$
 where $n$ is  another integer.   So in special case $r^2_{+} \sim n/m$ we
 have  a loophole. Assuming the "main" spectrum corresponds to
 winding number $m =1$ (the only stable state - all others with $m > 1$
 can decay into  states  with $ m = 1$)  we get the discrete spectrum
 $r_{+}^{2} \sim n$, i.e. the same spectrum $M \sim M_{p}\sqrt{n}$

Thus we see that the same  mass spectrum can be obtained from two
 absolutely different approaches. It will be interesting to understand
 if they will lead to equivalent predicitions in other cases. In the
 next section we shall consider  $2+1$-dimensional black holes
 and will find that one again will have $r_{+}^{2} \sim n$.

\section{Quantum Spectrum of 2+1 Black Holes.}

 Let us consider
 $2+1$ Einstein gravity coupled to abelian  topologically
 massive gauge field  defined by action
\bq
S = \int d^{3}x\left\{\frac{1}{\kappa} \sqrt{-g}\left[R  +  2l^{-2}\right]
 \frac{1}{2}\sqrt{-g}F_{\mu\nu}F^{\mu\nu}
 - m\epsilon^{\mu\nu\lambda}F_{\mu\nu}A_{\lambda} +
 \sqrt{-g}J^{\mu}A_{\mu}\right\}
\eq
 where $l^{-2}$ is the cosmological constant, $\kappa = 1/16\pi G$ is the
Planck mass and
 $J^{\mu}$ is the covariantly conserved current: $D_{\mu}J^{\mu}=0$.
  The coupled Einstein-Maxwell equations are
\bq
R_{\mu\nu} - \frac{1}{2} g_{\mu\nu} R = \kappa T_{\mu\nu}  +
g_{\mu\nu}l^{-2}\nonumber \\
\partial_{\nu}(\sqrt{-g}F^{\nu\sigma}) + m\epsilon^{\sigma\mu\nu}F_{\mu\nu}
 = \sqrt{-g}J^{\sigma}
\eq
where the stress-energy tensor $T_{\mu\nu}$ does not depend on the
 gauge Chern-Simons term and equals to
$T_{\mu\nu} = - F_{\mu\rho}F_{\nu}^{\rho} + \frac{1}{4}g_{\mu\nu}
F_{\lambda\sigma}F^{\lambda\sigma}$

We shall consider  solutions which depend only on radial coordinate $r$ only,
then the metric can be represented as
\bq
ds^{2}  = - N^{2}(r)dt^{2} + dr^{2} + R^{2}(r)d\theta^{2}
\label{metric}
\eq
 and only non-zero $F_{\mu\nu}$ components are  electric $F_{0r} = E(r)$
 and magnetic $F_{r\theta} = H(r)$ fields. It is easy to see that from
 $D_{\mu}J^{\mu} = 0$ one gets $J^{r} = 0$.
After simple calculations one gets ($X' = dX/dr$):
\bq
R_{00} & =  & NN'' + N N' \frac{R'}{R}=\kappa \frac{N^{2}}{R^{2}}H^{2}  +
2l^{-2}N^2
 \nonumber \\
R_{rr} & =  & -\frac{N''}{N} - \frac{R''}{R}=  - 2l^{-2} \\
R_{\theta\theta}  & = & - R R'' -  R R'\frac{N'}{N}=
 \kappa\frac{R^{2}}{N^{2}}E^{2}  -  2 l^{-2}R^{2}
 \nonumber
\label{1}
\eq
and
\bq
\frac{d}{dr}(\frac{R}{N} E) + mH = R NJ^{0}, ~~~~~~~~
\frac{d}{dr}(\frac{N}{R} H) + mE = R NJ^{\theta}
\label{2}
\eq
It is easy to see from (\ref{1}), (\ref{2})
 that Eistein-Maxwell equations are symmetric under
 the transformation
\bq
N \leftrightarrow R, \;\; E \leftrightarrow H, \;\;
J^{\theta} \leftrightarrow J^{0},\;\; \kappa \rightarrow
 -\kappa
\eq
which is easy to understand because of the formal
 symmetry between $\theta$ and $it$ in (\ref{metric}).

However, one can  put $J^{\theta} =0$ and get two solutions  with
 $E \sim N/R,\; H=0$ and $H \sim R/N,\; E=0$ considering point-like
 charge in pure Maxwell theory, i.e. Chern-Simons mass term is zero,
 $m=0$ or uniform  charge distribution in the topologically massive
 gauge theory with non-zero Chern-Simons term
 $m \neq 0$. Then we see that the abovementioned
 duality can be realised not as duality betwen time and angular components
 of the current in the same theory, but as a duality between
 point-like charge distribution in pure Maxwell theory and uniform
 charge distribution in topologically massive gauge theory.

In \cite{btz} the case $m = 0,~ l^{-2} \neq 0$  has been considered and in
 \cite{kogan2} the opposite case case with zero cosmological constant $l^{-2} =
0$
and nonzero $m$.   Let us consider the neutral black hole \cite{btz}.
\bq
 \frac{N''}{N} + \frac{N}{ N'} \frac{R'}{R}  = 2l^{-2}, ~~~
\frac{N''}{N} + \frac{R''}{R}  =  2l^{-2},~~~~
 \frac{R''}{R} + \frac{R'}{R}\frac{N'}{N} =
  2l^{-2}
\label{1E}
\eq
 from which  one gets
\bq
  \frac{R''}{R} = l^{-2}, ~~~ N  = R'
\eq
The black hole metric  can be rewritten in the form  \cite{btz}
\bq
ds^2 = -N^{2} dt^2 + \frac{dR^2}{N^{2}} + R^{2}d\theta^{2}
\eq
where
\bq
R = R_{+} \cosh (r/l), ~~~ N = R' =\sqrt{(R^{2} - R_{+}^{2})l^{2}}
\eq
and $R_{+}$ is the  radius of the horison. The mass of the black hole
 (in Planck units) is $M = (R_{+}/l)^{2}$ and  one can find Hawking temperature
 using the analytical continuation $t \rightarrow it$. Then to have regular
 manifold with $R \leq R_{+}$ one has to  make  $t$ periodic variable
with period  $2\pi l^2/R_{+}$ and the temperature is $T = R_{+}/2\pi l^{2} =
 \sqrt{M}/2\pi l$. The manifold has topologu $R^{2} \times S^{1}$, where
 $S^{1}$ corresponds to angle $\theta$.

 Using the same arguments about decoupling of left- and right moving modes
  (but in this case for angle coordinate $\theta$, not $t$) one can conclude
 that quantization condition must be $R_{+}^{2} \sim n$ and thus mass
 must be quantized as $M \sim n$, not $\sqrt{n}$. Moreover spectrum
 for $R_{+}^{2}$ (but not for mass $M$)
must be independent on cosmological constant $l^{_2}$.

It is amusing that this argument is supported by  Carlip results \cite{carlip}
who  used a Chern-Simons \cite{cs1}, \cite{cs2} description for the
 $2+1$ gravity. Using the fact the horison acts as a boundary he got
 an effective  boundary dynamics describing the dynamical degrees of freedom
 on horison (for details see his paper) and found that the entropy of black
hole
 $S = 2\pi R_{+}/4G$ is equal to the logarithm of number of  boundary states.
 To be more precise, he found that the Virasoro operator on the boundary is
\bq
L_{0} = N - (R_{+}/4G)^{2}
\eq
where $N$ is a number operator. Using the fact that the number of states for
 given $N$ behaves as $n(N) \sim exp(\pi\sqrt{4N})$ he found that entropy
\bq
S = ln n(N) = 2\pi \sqrt{N} = 2\pi R_{+}/4G
\eq
which is the correct expression for the entropy.

However, this means at the same time that Virasoro constarint
\bq
L_{0} = 0
\eq
means the quantization condition for the black hole
\bq
\frac{R_{+}^2}{16 G^{2}} = N
\eq
- the same as one can get from the abovementioned argument. Let us
 also note that there is no dependence on $l^{-2}$ - which means
 that this quantum spectrum can not be obtained from any purely
 thermodynamical arguments (because both mass and tempearture
 depends explicitly on cosmological constant).

One can also consider the case of  the black holes interacting with
 the TMGT. IN this case one can get additional contribution
 to the Virasoro operator due to the induced gauge degrees of freedom
 at the horison. For abelian Chern-Simons theory this contribution
 to the Virasoro generator will be proportional to $A_{\theta}^{2}$
 and using the solution with the uniform density of charge which has
 been considered in \cite{kogan2} one can get the spectrum in this case.

\section{Conclusion}

 In this letter we tried  to compare some handwaving arguments in
 favour of discrete spectrum of black holes and demonstrate that
 at least in one case - $2+1$ dimensional black holes - one
 can obtain this spectrum.  It will be  extremely interesting
 to understand if the same spectrum can be reproduced in  membrane
 approach advocated by Maggiore \cite{maggiore1}, \cite{maggiore2} -
 and if there is any connection between spectral condiiton in his
 approach (existence of normalizable wave function for fluctuating
 membrane) and existence of chiral  string modes.

\end{document}